\documentclass[conference]{IEEEtran}
\IEEEoverridecommandlockouts
\usepackage{cite}
\usepackage{amsmath,amssymb,amsfonts}
\usepackage{algorithmic}
\usepackage{graphicx}
\usepackage{textcomp}
\usepackage{xcolor}
\usepackage[utf8]{inputenc}
\usepackage{tikz}
\usepackage[table]{xcolor}
\usepackage{xspace}
\usepackage{multirow}
\usepackage{tabularx}
\usepackage{booktabs}
\usepackage{graphicx}
\usepackage{caption}
\usepackage{wrapfig}
\usepackage{balance}
\usepackage{xurl}
\usepackage{rotating}
\usepackage{hyperref}
\usepackage{flushend}
\usetikzlibrary{positioning,shapes.geometric,shapes.misc}
\usepackage[normalem]{ulem}
\usepackage{enumitem}
\usepackage{subcaption}
\usepackage[most]{tcolorbox} % <-- needed for \finding
\usepackage{makecell}
\usepackage{caption} 
\usepackage{caption} 
\newcolumntype{L}{>{\raggedright\arraybackslash}X} % left-aligned, stretches
\newcolumntype{R}{>{\raggedleft\arraybackslash}X}  % right-aligned, stretches
\newcolumntype{C}{>{\centering\arraybackslash}X}   % centered, stretches

\definecolor{codegreen}{rgb}{0,0.6,0}
\definecolor{codegray}{rgb}{0.5,0.5,0.5}
\definecolor{codepurple}{rgb}{0.58,0,0.82}
\definecolor{backcolour}{rgb}{0.95,0.95,0.92}
\definecolor{medium-blue}{rgb}{0,0,1}

\hypersetup{
  colorlinks=true,
  linkcolor=black,
  urlcolor=medium-blue,
  citecolor=black
}

\newcommand{\name}{\textit{Reprodgen}\xspace}
\newcommand{\namebenchbase}{\textit{ReprodgenBench-B}\xspace}
\newcommand{\namebenchextend}{\textit{ReprodgenBench-E}\xspace}
\newcommand{\namebenchverified}{\textit{ReprodgenBench-V}\xspace}
\newcommand{\namebenchissues}{\textit{ReprodgenBench-GI}\xspace}

\newcounter{NumObservations}
\newcommand{\finding}[1]{%
    \vspace{-0.5em}
  \begin{tcolorbox}[
      enhanced,
      breakable,
      colback=gray!5,
      colframe=black,
      boxrule=0.5pt,
      arc=2pt,
      left=6pt,
      right=6pt,
      top=4pt,
      bottom=4pt,
      width=\linewidth
  ]
  \textbf{Finding \arabic{NumObservations}:}~#1
  \end{tcolorbox}
  \vspace{-0.5em}
  \stepcounter{NumObservations}%
}

\usetikzlibrary{positioning,shapes.geometric,shapes.misc}

\newcommand*\circleB[1]{%
  \tikz[baseline=(char.base)]{
    \node[
      shape=circle,
      fill=gray!15!white,
      inner sep=1pt
    ] (char) {\textcolor{black}{#1}};
  }%
}

\newcommand{\circleT}[1]{%
  \raisebox{0.2pt}{\textcircled{\raisebox{-0.3pt}{\small #1}}}%
}

\newtcolorbox{examplebox}[1][]{
  enhanced,
  breakable,
  colback=gray!5,
  colframe=gray!50,
  boxrule=0.4pt,
  arc=2pt,
  left=6pt,
  right=6pt,
  top=4pt,
  bottom=4pt,
  width=\linewidth,
  fontupper=\small,
  title={#1},
  fonttitle=\bfseries,
  coltitle=black,
  halign title=center
}

\def\BibTeX{{\rm B\kern-.05em{\sc i\kern-.025em b}\kern-.08em
    T\kern-.1667em\lower.7ex\hbox{E}\kern-.125emX}}
\begin{document}

\title{From Discussion to Execution: Replicating Buggy and Correct Data Science Code}

\author{\IEEEauthorblockN{Ragib Shahariar Ayon}
\IEEEauthorblockA{\textit{Dept. of Computer Science} \\
\textit{Texas State University}\\
San Marcos, Texas \\
ipd21@txstate.edu}
\and
\IEEEauthorblockN{Mohammad Wardat}
\IEEEauthorblockA{\textit{Dept. of Computer Science and Engineering} \\
\textit{Oakland University}\\
Rochester , MI \\
wardat@oakland.edu}
\and
\IEEEauthorblockN{Shibbir Ahmed}
\IEEEauthorblockA{\textit{Dept. of Computer Science} \\
\textit{Texas State University}\\
San Marcos, Texas \\
shibbir@txstate.edu}
}

\maketitle
\thispagestyle{plain}
\pagestyle{plain}

\begin{abstract}
\label{sec:abstract} 
Reproducing reliable data science code from informal sources is challenging due to ambiguous problem specifications, missing dependencies, and performance bottlenecks. Although developer Q\&A forums provide rich discussions on diagnosing and fixing real-world issues, the information is often incomplete and unstructured, limiting its use for automated debugging and verification. In this paper, we introduce \name, a large language model (LLM) based framework for automatically replicating executable buggy and patched data science programs from Q\&A forum posts. Given a question and its corresponding answer, \name reconstructs the buggy behavior described in the question and the intended fix described in the answer, producing executable buggy–patched code pairs that reflect the original discussion. The framework builds structured representations of code intent (CI), functional requirements (FR), and Structured Chain of Thought (SCoT), and iteratively refines code using an LLM-based reviewer until it is executable and semantically consistent. We evaluate \name on Stack Overflow (SO) and GitHub Issues (GI) across seven data science libraries, including pandas, numpy, and scikit-learn, and construct a benchmark of runnable buggy–patched programs validated by human experts. Our pipeline uses LLMs for semantic assessment, while executability is verified through actual execution. Results show reliable replication with clear differences in model performance.

\end{abstract}


\section{Introduction}
\label{sec:introduction}
\begin{figure*}[t]
%\begin{wrapfigure}{R}{5cm}
\centering
	\includegraphics[width=\linewidth,trim={0cm 0cm 0cm 0cm},clip]{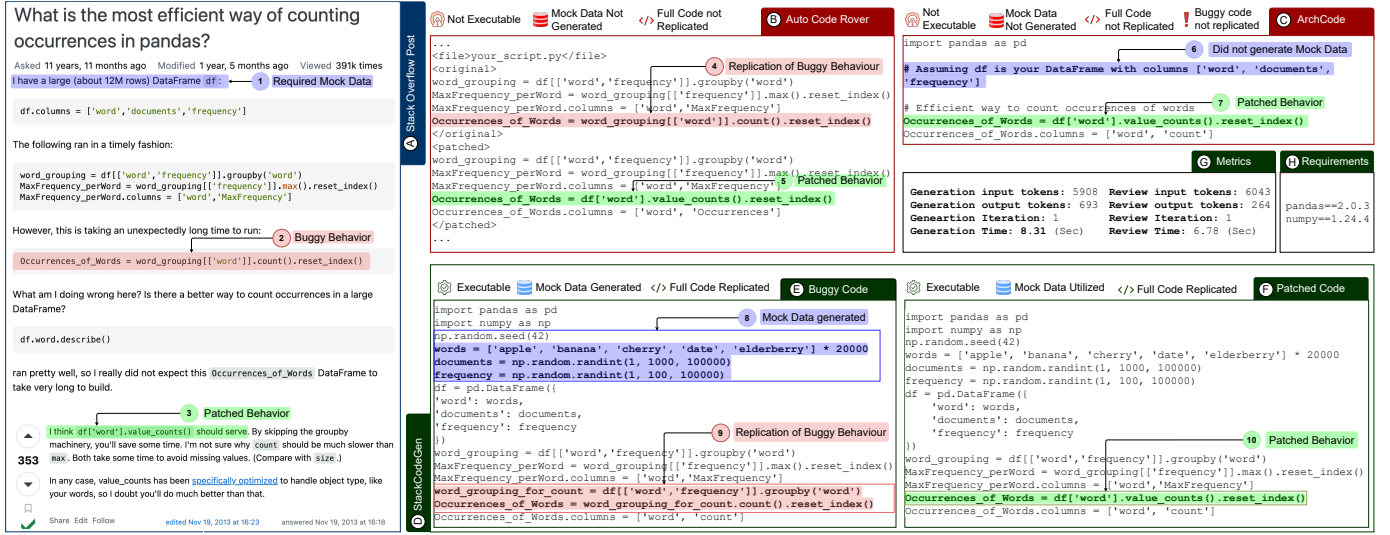}
	\footnotesize
	\caption{Motivating example illustrating how \circleT{D} \name generates fully executable \circleT{E} buggy and \circleT{F} patched code, along with \circleT{G} metrics and \circleT{H} inferred \texttt{requirements.txt}, from \circleT{A} Question and Answer forum post \#20076195. In contrast, prior state-of-the-art (SOTA) approaches such as \circleT{B} AutoCodeRover and \circleT{C} ArchCode typically produce partial, non-executable snippets that fail to replicate the complete functionality of the original code.}
	\label{fig:motiv}
%\end{wrapfigure}
\end{figure*}
Debugging data science programs is essential for producing trustworthy and accurate outcomes, particularly when dealing with subtle issues such as performance bottlenecks, logical mistakes, or computational errors~\cite{Islam2019FSE, Islam2020ICSE, ahmed2023characterizing, yang2025towards, chen2025towards}. Researchers often rely on curated benchmarks containing buggy and corrected code samples to evaluate debugging, repair, and verification techniques in a controlled and reproducible manner~\cite{Islam2019FSE, yang2025towards, chen2025towards, SantanaTOSEM, wang2025machine}. However, constructing such benchmarks from informal developer Q\&A posts remains difficult. These posts frequently omit crucial details, such as library dependencies, execution context, functional requirements, and example datasets, making it challenging to reconstruct an executable environment. Reproducing the buggy behavior described in these posts is a prerequisite for reliable debugging and automated patch generation. Executable buggy code enables systems to analyze runtime behavior, validate inferred requirements, and verify whether a generated patch actually resolves the defect. Moreover, reproducible buggy--patched pairs form the basis for benchmarking automated repair systems and studying the reasoning capabilities of LLMs in realistic debugging scenarios. Without executable buggy code, it is difficult to measure correctness, performance improvement, or semantic equivalence between faulty and fixed versions, and therefore difficult to faithfully reproduce and repair the bugs discussed in such posts~\cite{Islam2019FSE, Islam2020ICSE, yang2025towards}.

Existing empirical studies have examined a wide range of debugging and fault-localization issues in data science programs. Prior work has analyzed common coding errors in Jupyter Notebooks and in Python or R scripts~\cite{Islam2019FSE, ahmed2023characterizing, zhao2024vulnerableR, chen2025towards}, as well as performance, logical, and semantic issues in widely used data science libraries such as NumPy, pandas, Matplotlib, and scikit-learn~\cite{yang2025towards}. In parallel, recent studies have investigated how effectively LLMs can generate code and assess patch correctness~\cite{ouyang2025knowledge, ruan2025specrover, nascimento2025effective, molina2024improving, zhao2024llmsR, WangJudgeLLM}. Despite this progress, a key gap remains: the automated reconstruction of executable buggy--patched code pairs directly from informal sources such as developer Q\&A forum posts. Addressing this problem requires more than producing syntactically valid code; it also requires semantic faithfulness so that the generated buggy code reproduces the intended faulty behavior and the patched code reflects the corresponding fix~\cite{ouyang2025knowledge, fei2025patch, zhou2023patchzero, yang2025towards}.

To address this gap, we propose \name, an LLM-based framework for generating executable buggy and patched Python code from developer Q\&A posts, including Stack Overflow discussions and GitHub Issues. The framework follows a Generator--Reviewer workflow in the Driver--Navigator style of pair programming. The Generator LLM produces intermediate artifacts and candidate code, while the Reviewer LLM checks whether these outputs remain aligned with the source post. In this workflow, \name derives CI, FR, and SCoT from the question and answer, uses them to guide code synthesis, and then evaluates the generated buggy and patched programs using execution signals such as standard output, error stream, and exit code. 

Designing \name required addressing several technical challenges that arise when reconstructing executable buggy--patched pairs from informal Q\&A discussions. Source posts often omit essential execution context, requiring the framework to infer missing dependencies, synthesize mock data, and fill in missing code elements. The framework must also preserve semantic consistency across multiple stages of generation, since errors introduced in early artifacts can propagate to later code synthesis. To address this, \name combines iterative Generator--Reviewer refinement with staged validation, so that intermediate artifacts are reviewed and revised before they are consumed downstream, helping limit error accumulation during buggy and patched code generation.

As part of this work, we also constructed \namebenchverified, a benchmark of 176 real-world Python programs derived from developer forum discussions spanning seven widely used data science-related Stack Overflow tags (NumPy, pandas, DataFrame, Matplotlib, scikit-learn, SciPy, and PyQt)~\cite{soTagtrends}. For each instance, a single executable buggy--patched program pair was selected from LLM-generated candidates and independently validated by two expert human judges to ensure correctness and faithfulness to the original discussion. On the base dataset, \namebenchbase, our approach, when paired with Qwen~2.5 and intermediate artifacts, achieved execution rates of 60.00\% for buggy code and 52.21\% for patched code, and corresponding success rates of 58.12\% and 51.72\%, respectively, with an average generation time of 35~seconds per issue. In addition, this work also contributes to the following:
\begin{itemize}
 \item Introducing \name, a pair-programming framework leveraging LLMs to generate and review executable buggy and patched code from informal Stack Overflow posts.

\item Developing a structured prompt-engineering approach combining code intent, functional requirement, structured chain-of-thought, utilizing iterative assessment feedback to ensure semantic clarity and correctness.

\item Creating and publicly releasing \namebenchverified, a benchmark consisting of 176 real-world data-science programs curated from Stack Overflow related to seven widely used data science-related tags, promoting reproducibility and further research.

\item Presenting a community-driven leaderboard for the \name~benchmark that reports key metrics such as Execution Rate, Success Rate, and semantic similarity, enabling transparent model comparison, reproducibility, and continuous community contributions to automated code generation and debugging research.
\end{itemize}

\section{Motivation}
\label{sec:motivation}
Constructing executable code that contains both the original bug and its corresponding patch from informal descriptions such as Stack Overflow posts~\cite{stackoverflow:20076195} is challenging. These posts often lack essential details, such as representative datasets, instructions, or dependencies, making reproduction or solution validation difficult. Although LLMs can generate code from textual descriptions, unconstrained generation often yields programs that are syntactically valid but semantically incorrect, either by misinterpreting user intent or introducing new bugs.

\begin{figure*}[!t]
\centering
\includegraphics[width=\linewidth,trim={0cm 0cm 0cm 0cm},clip]{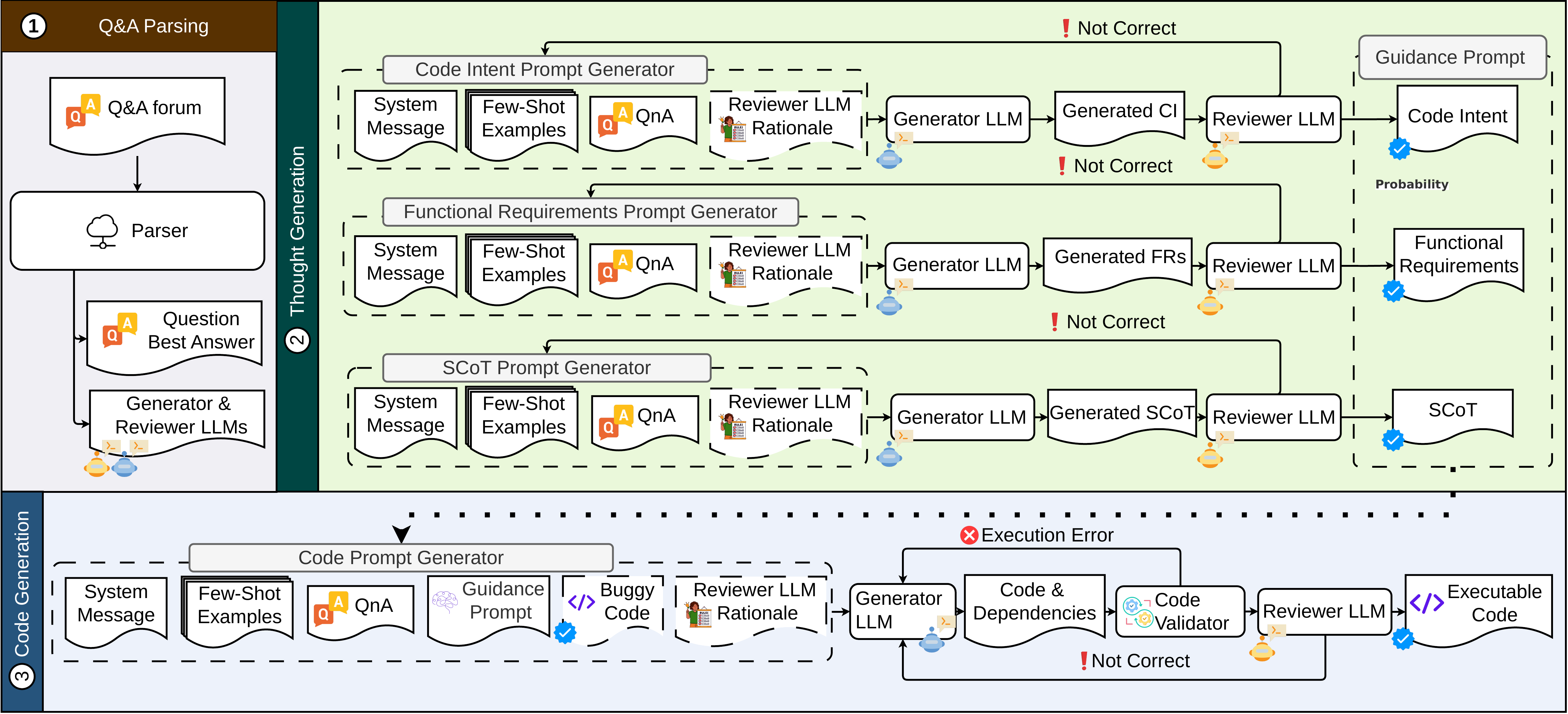}
\vspace{-0.5em} % adjust this value as needed
\captionsetup{font=footnotesize}
\caption{Overview of \name framework}
\label{fig:overview}
\end{figure*}

To illustrate the problem, let us consider a real Stack Overflow example~\cite{stackoverflow:20076195}, as shown in Figure~\ref{fig:motiv}, which demonstrates this challenge. \circleB{1} The post discusses counting word occurrences in a large \texttt{pandas} DataFrame (about 12M rows). Users of Stack Overflow describe the problem in natural language texts. It mentions dataset size and runtime issues, but does not contain fully executable code, example data, dependencies, or environment details. The accepted answer replaces the inefficient \circleB{2} \texttt{df.groupby('word').count()} with \circleB{3} \texttt{df['word'].value\_counts()}. Automatically reproducing this requires understanding the user's intent, generating mock data, and ensuring consistent execution. Typical LLM-generated code mimics inefficient logic but fails to capture intended optimizations or necessary context. 

To evaluate whether existing state-of-the-art approaches can reproduce both the reproducible buggy code and its corresponding patched version, we experimented with AutoCodeRover and ArchCode, investigating how they performed for some posts~\cite{ReprodgenSota}. AutoCodeRover (ACR) combines LLMs with code search to resolve GitHub issues~\cite{AutoCodeRover2024}. To compare accurately, we recreated the Stack Overflow post as a GitHub issue, using its title and question body. After that, we were able to execute ACR in \textit{GitHub Issue Mode}. However, ACR assumes a complete repository context with source files and dependencies. Because Stack Overflow posts lack this information, context retrieval failed in such cases. ACR generated a textual patch transforming \circleB{4} \texttt{groupby().count()} to \circleB{5} \texttt{value\_counts()}. Although it semantically matches the answer, it is missing mock data, imports, or dependencies. The output was non-executable and unverifiable, highlighting ACR’s limitations for Q\&A-style tasks. Another work, ArchCode, generates executable code from textual requirements~\cite{han2024archcode}. Given the same post, it misinterpreted the question "Is there a better way to count occurrences?" as an instruction to produce the optimized solution rather than the buggy code. Consequently, it generated \circleB{7} \texttt{value\_counts()} instead of \texttt{groupby().count()}. Moreover, \circleB{6} ArchCode assumed an existing populated \texttt{DataFrame}, producing no mock data or dependencies. The generated output was again non-executable for both SoTA AutoCodeRover and ArchCode.

To overcome these issues, we developed \name, a structured framework that validates each generation step. \name first generates and evaluates CI, FR, and ScoT. Then another LLM reviews to detect semantic errors. After that, it iteratively refines outputs until the final code aligns with the intended behavior. For buggy code, \name can successfully generate mock data, dependencies, and fixed random seeds (\circleB{8}, \circleB{9}) to ensure exact reproducibility mentioned in the Stack Overflow posts. For correct code or patched code generation, it uses accepted answers to produce the optimized \circleB{10} \texttt{value\_counts()} version. The resulting buggy-patched pair is executable and semantically reliable, even when the original problem lacks full details in the Stack Overflow posts. These observations reveal a critical research gap: existing techniques either generate functional code without reproducing the buggy code (ArchCode) or produce non-executable patches (AutoCodeRover). \textbf{To fill this gap, we introduce \name, a framework that reliably generates executable, buggy-patched pairs from informal text, such as Stack Overflow posts or GitHub Issues.}

Unlike prior systems, \name integrates validation and refinement in its pipeline. It generates intermediate artifacts while generating executable buggy and patched code: \textit{CI}, \textit{FR}, and \textit{SCoT}, which a \textit{Reviewer LLM} examines for inconsistencies and refines iteratively. While generating buggy code, \name simulates missing context via mock data, dependency inference, and fixed seeds, producing an executable version of the buggy code \circleB{9}. For patches, it synthesizes corrected code reflecting expert-provided fixes \circleB{10} in the answer. The resulting code pairs are executable, semantically consistent, and reproducible. Overall, \name bridges key gaps by unifying mock data synthesis, dependency inference, semantic validation, and executability, thereby making it directly applicable to Q\&A-style problem descriptions that lack complete code context.

\section{Approach} 
\label{sec:approach}
\textit{Problem description:} Given a developer Q\&A post $\mathcal{P}$, the goal is to automatically generate buggy code $\mathcal{C}_b$ and patched code $\mathcal{C}_p$, such that $\mathcal{C}_b$ reflects the user’s buggy behavior described in the question, and $\mathcal{C}_p$ implements the solution described in the corresponding answer. Such posts commonly appear in platforms like Stack Overflow or GitHub issue discussions, where users describe a problem, and others provide fixes or workarounds.

\subsection{Overview}
\label{sec:Overview}Figure~\ref{fig:overview} presents the end-to-end workflow of \name, from parsing a developer Q\&A post to producing executable buggy and patched code. Given a post identifier, \name first applies a Q\&A parser \circleB{1} to extract the question, a selected answer, and associated metadata. A generator--reviewer LLM pair is then selected, either user-specified or from a predefined ranked list. Next, \name performs a shared thought generation stage \circleB{2} for both buggy and patched code, producing structured intermediate guidance artifacts that assist code synthesis. In this paper, we use the terms \emph{guidance generation} and \emph{thought generation} interchangeably. Importantly, this stage is optional: \name can directly proceed to code generation without it, enabling a lightweight baseline configuration. This design makes the framework modular, as the guidance component can be included, removed, or replaced without affecting the overall pipeline. Finally, in the code generation stage \circleB{3}, \name generates executable buggy and patched programs, validates them through execution, and refines them until they are both runnable and semantically faithful to the input post. The prompt templates~\cite{ReprodGenPrompts}, generated thought processes~\cite{ReprodgenResults}, and their corresponding review rationales for the example discussed in Section~\ref{sec:motivation} are available in the repository.

\subsection{Q\&A Parsing}
The Q\&A parser processes developer posts consisting of a question and associated responses, obtained from forums or software repositories. It uses platform-specific APIs or structured interfaces to retrieve the question content, candidate answers, and relevant metadata. Given a post identifier, the parser extracts the question (e.g., title and body) along with all available responses (e.g., answers or discussion comments). Since multiple responses may exist, the system selects a single answer for downstream processing. If an explicitly marked preferred answer is available (e.g., an accepted answer), it is selected by default; otherwise, the response with the highest relevance signal (e.g., score or community feedback) is chosen. In cases of ambiguity, the final selection is determined through manual inspection to ensure consistency and relevance. The extracted question and selected answer are then used as input to the code generation pipeline.

\subsection{In-Context Learning: Thought generation}
\label{thought_generator} 
To replicate the buggy--patched code pairs, we first retrieve relevant context from each Q\&A post. We then extract three forms of semantic guidance from the question and the selected answer: \textit{CI}, \textit{FR}, and a \textit{SCoT}. These artifacts guide code synthesis and are iteratively refined using feedback from an independent reviewer LLM to ensure accuracy and completeness.

\subsubsection{Code Intent (CI)}
CI serves as a semantic anchor for both buggy and patched code generation, ensuring that synthesized programs preserve the same underlying intent while differing only in correctness. Inspired by function summary generation in SpecRover~\cite{ruan2025specrover}, we extract CI as a high-level specification of the developer’s intended functionality. We distinguish two forms: \textit{Buggy CI}, which reflects the intended goal described in the question, and \textit{Patched CI}, which captures the intended behavior after applying the selected fix.

\subsubsection{Functional Requirements (FR)}
We extract FR as structured, user-oriented specifications derived from the Q\&A post. Inspired by requirement synthesis in ArchCode~\cite{han2024archcode}, FRs describe expected behavior, input-output relationships, and relevant edge cases, independent of implementation details. We construct two variants: \textit{Buggy FRs}, which reflect the user’s original (possibly incomplete or incorrect) expectations, and \textit{Patched FRs}, which capture the corrected functionality described in the answer. Each FR is generated in a structured format and reviewed by a separate LLM judge. Detected inaccuracies are iteratively refined using reviewer feedback. These validated FRs serve as executable specifications guiding code generation.

\subsubsection{Structured Chain of Thought (SCoT)}
We extract SCoT representations to capture the procedural reasoning underlying the code. Inspired by structured reasoning approaches~\cite{li2025structured, wei2022chain}, SCoT provides a step-by-step description of how inputs are transformed into outputs, reflecting control flow while abstracting implementation details. We construct two variants: \textit{Buggy SCoT}, which models the user’s reasoning leading to incorrect behavior, and \textit{Patched SCoT}, which captures the corrected procedural logic described in the answer. Each SCoT includes an input-output signature and an ordered sequence of reasoning steps. These artifacts are independently reviewed and refined to ensure completeness and faithfulness.

\subsection{Code Generation}
\name generates executable buggy and patched code through an iterative generate-and-execute loop. Given a Q\&A post and optional validated guidance artifacts, the system synthesizes runnable code along with inferred dependencies and mock inputs. Each generated artifact is executed to check whether it runs successfully. Execution failures are used to refine subsequent generations until the artifact becomes executable or a retry limit is reached. After execution, the reviewer LLM evaluates whether the generated program behavior, including its standard output and error stream, faithfully reflects the intended buggy behavior or patched behavior described in the input post.

\textit{Code Executor:}
\label{code-executor}
To ensure safe and reproducible execution, we implement a \texttt{Code Executor} that runs generated programs in a separate, isolated execution environment with its own runtime and dependencies. For each artifact, the executor creates a temporary workspace, writes the generated Python script and dependency specification, and constructs the environment using the inferred Python version. Dependencies are installed within this isolated environment, after which the script is executed. The executor captures the standard output, error stream, and exit code to determine executability. If execution fails, the resulting error feedback is passed back to the generator LLM to guide refinement.

\subsection{Reviewer LLM}
\label{code-validator}
The reviewer LLM evaluates intermediate outputs from both the thought generator and code generator. Following prior work~\cite{le2019reliability}, we use four labels: \textit{Correct}, \textit{Partially Correct}, \textit{Incorrect}, and \textit{Unknown}, along with a rationale. For guidance artifacts, the reviewer is provided with the triplet of CI, FR, and SCoT, and verifies whether they are grounded in the original question and answer, checking for hallucinations~\cite{zhang2025llm} and ensuring coverage of all relevant behaviors. To review generated code, the reviewer LLM receives the original Q\&A post, the generated code, and its execution results, including standard output, error stream, and exit code. It then judges whether the executed behavior faithfully reflects the intended buggy behavior or the corresponding fix described in the question and selected answer. The label and rationale generated by the reviewer LLM are used as structured feedback for the thought generator and code generator LLMs, supporting iterative refinement toward executable and semantically faithful buggy and patched code pairs. In this paper, we use the terms \textit{Reviewer LLM} and \textit{Judge LLM} interchangeably.
\section{Evaluation}
\label{sec:evaluation}
\subsection{Experimental setup}

\subsubsection{Research Questions}
Our evaluation focuses on three research questions.
\begin{itemize}[leftmargin=*]
    \item \textbf{RQ1 (Effectiveness):} How effective is~\name in replicating buggy and patch executable codes from informal issue descriptions?
    \item \textbf{RQ2 (Reliability):} How reliable is the generated buggy and patched code that reproduces the bug and applies the patches?
    \item \textbf{RQ3 (Efficiency):} What is the time and budget to generate and assess the patches, and the number of iterations? 

\end{itemize}

\subsubsection{Dataset}
To identify candidate posts, we queried the top 500 Stack Overflow posts for data-science-related tags: NumPy, Pandas, Matplotlib, scikit-learn, SciPy, DataFrame, and PyQt, with a score of at least 10, using a publicly available Stack Exchange query tool~\cite{StackExchangeQuery,soTagtrends}. During collection, 18 posts could not be parsed or failed to meet the score requirement, yielding 3,482 successfully parsed posts prior to filtering. We then applied heuristic-based filters to remove posts without accepted answers, without code blocks, or focused on crash-related issues, which eliminated 1,477 posts and left 2,005 candidates. Next, we applied a conservative LLM-based semantic filter to distinguish bug-related posts from non-bug-related discussions, identifying 578 candidates that plausibly described buggy behavior. After de-duplication, two annotators independently reviewed these posts and further rejected 100 cases that did not involve genuine buggy behavior or were not sufficiently challenging. The annotators reached consensus on 427 Stack Overflow posts, forming \namebenchextend. 

\begin{figure}[!htbp]
    \centering
    \includegraphics[width=1\columnwidth]{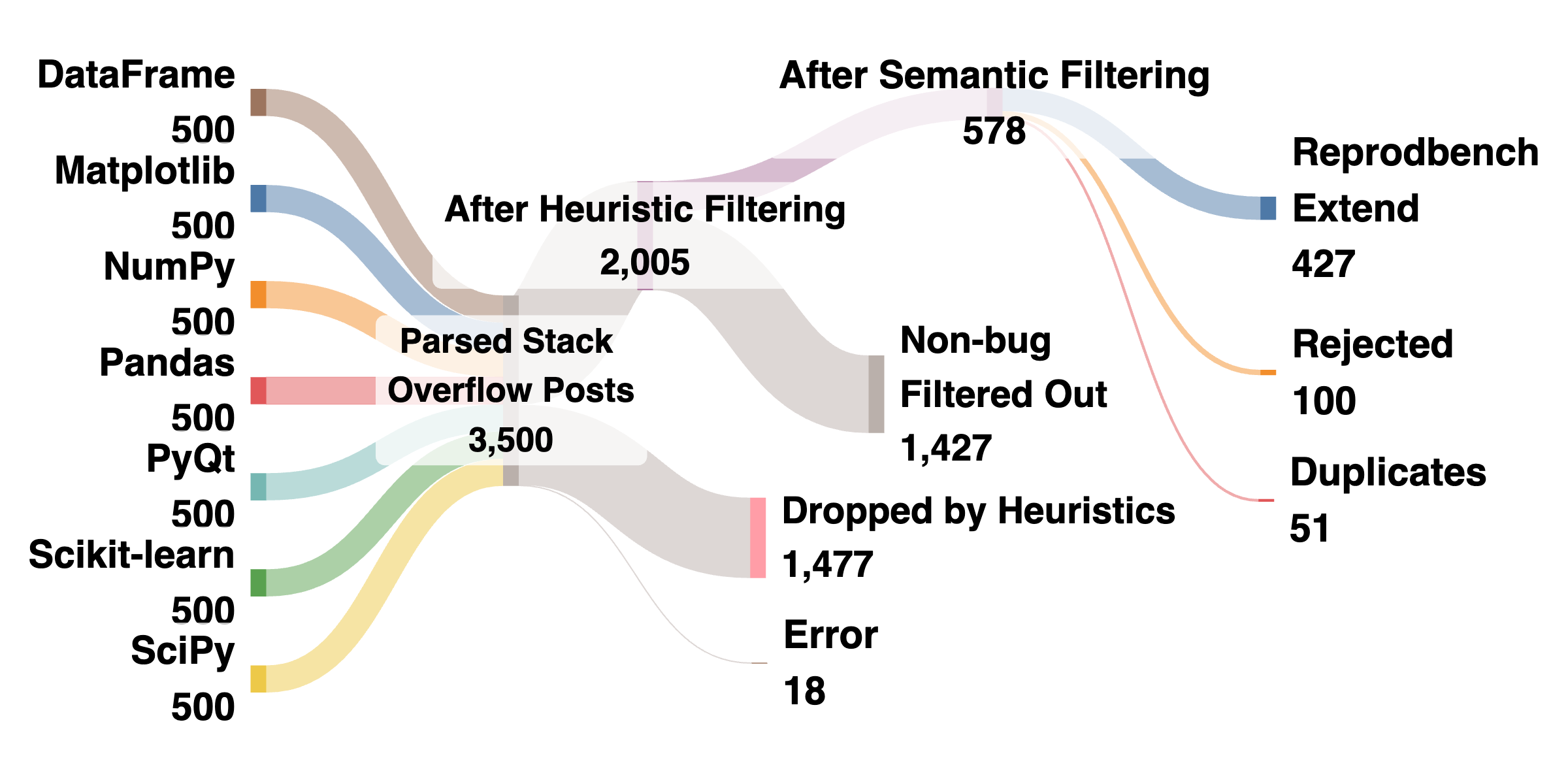}
    \caption{Sampling methodology for \namebenchextend.}
    \label{fig:placeholder}
\end{figure}

\paragraph{{\namebenchbase}}
Following the sampling methodology of Misu et al.~\cite{10.1145/3643763} and to stay within budget constraints, we randomly sampled 203 posts from this curated set using a 95\% confidence level and a 5\% margin of error, yielding \namebenchbase, which is used for all experimental evaluations for Stack Overflow in this study. All artifacts in \namebenchbase are first synthesized using an LLM pair, Claude Sonnet 4.5 as generator and GPT-4o as reviewer, as candidate outputs. These synthesized artifacts are not treated as ground truth. Instead, two expert judges with over five years of debugging research experience independently assessed all synthesized artifacts using four labels: \textit{Correct}, \textit{Partially Correct}, \textit{Incorrect}, and \textit{Unknown}~\cite{zhang2025llm}. All final ground-truth labels are determined solely by the human judges. To avoid evaluation bias, the synthesis language models are not used at any stage of evaluation or judgment. Following established annotation protocols that continuously monitor inter-annotator agreement using Cohen’s~$\kappa$~\cite{10.1145/3338906.3338955}, labeling was conducted in multiple seating windows. In the first window of 25 posts, agreement was low for buggy-code labels and moderate for patched-code labels ($\kappa=0.32$), reflecting the inherent ambiguity of reproducing buggy behavior from informal descriptions. We therefore conducted a calibration session to refine labeling guidelines using concrete examples. In the subsequent window, agreement converged to perfect for both buggy and patched code labels ($\kappa=1.0$). Disagreements from earlier windows were adjudicated by another author acting as mediator, after which all instances were assigned final consensus labels. Once a stable agreement was reached, no further reconciliation meetings were required during annotation of the full \namebenchbase dataset~\cite{biswas2022}.

\paragraph{{\namebenchissues}}
To evaluate the applicability of \name\ beyond Stack Overflow, we sample 10 GitHub Issues (closed) from each of five libraries: NumPy, Pandas, Matplotlib, scikit-learn, SciPy, following the same filtering criteria as \namebenchextend. Two candidate tags are excluded: \texttt{dataframe}, which does not correspond to a standalone repository, and \texttt{PyQt}, which is not publicly accessible. We use GitHub Issues as a complementary data source, as they capture real-world developer-reported problems in a discussion-based format. Although structurally different from Stack Overflow, they exhibit similar challenges, including incomplete context, implicit assumptions, and iterative clarification through comments. Unlike \namebenchbase, we do not rely on LLM synthesis for this dataset. Instead, one author manually reproduces each issue by executing the reported buggy code to validate the failure behavior, and identifies the resolution by selecting the comment or discussion thread that provides a confirmed fix or viable workaround. This process yields a curated set of 50 issues, referred to as \namebenchissues. Compared to \namebenchbase, which uses LLM-synthesized artifacts validated by human judges, \namebenchissues is fully human-validated, providing a complementary evaluation setting.

\subsubsection{Large Language Models}

In this work, we employed open-source LLMs for two primary tasks: \textit{code generation} and \textit{review}. To comprehensively assess the performance of open-source models, we incorporated a diverse set of state-of-the-art models spanning different families and parameter scales choosen from a publicly available leaderboard~\cite{bigcode-leaderboard}: Qwen 2.5-Coder (32B)~\cite{yang2025qwen3}, Gemma 3 (27B)~\cite{team2025gemma}, LLaMA 4 (16x17B)~\cite{meta2024llama3}, DeepSeek-Coder-V2 (16B)~\cite{zhu2024deepseek}, Phi-4 (14B)~\cite{abdin2024phi}, Qwen 3 (8B)~\cite{yang2025qwen3}, and Code Llama (7B). These models represent a broad spectrum of capabilities, ranging from general-purpose to code-specialized architectures. To reduce evaluation bias, the models alternated between generator and reviewer roles. 

\subsubsection{Evaluation metrics}
To evaluate the code generation process, we used Execution Rate (ER), Success Rate (SR), CodeBERTScore, and Levenshtein distance, and for evaluating the reviewer LLM, we compared the generated labels against human expert annotations and calculated the classification metrics: Precision, Recall, and F1 Score.

\textit{Execution Rate (ER):} 
ER measures the proportion of generated programs that execute successfully without runtime exceptions~\cite{yuan2024evaluating}. Programs that run to completion but produce incorrect outputs are still considered successful under this metric.
\[
ER = \frac{N_{\text{executed}}}{N_{\text{total}}}
\]
where $N_{\text{executed}}$ is the number of programs that execute successfully, and $N_{\text{total}}$ is the total number of generated programs. Thus, ER captures the executability of generated programs over the dataset.

\textit{Success Rate (SR):} 
SR measures the proportion of generated programs that successfully reproduce the target behavior, as determined by a judge function that compares the generated code with the expected behavior~\cite{chao2024jailbreakbench}.
\[
SR = \frac{1}{N_{\text{total}}} \sum_{i=1}^{N_{\text{total}}} \mathbf{1}\!\left[\textsc{Judge}(C_i, B_i) = \text{correct}\right]
\]
where $C_i$ is the generated program for instance $i$, $B_i$ is the expected target behavior, and $\mathbf{1}[\cdot]$ is the indicator function that returns 1 if the condition holds and 0 otherwise.

\textit{CodeBERTScore:} 
CodeBERTScore~\cite{zhou-etal-2023-codebertscore,10.1145/3180155.3180206} measures the semantic similarity between generated and reference code using contextual embeddings from a pretrained CodeBERT model (microsoft/codebert-base), following prior work. It computes token-level similarity by matching each token in one sequence to its most similar token in the other sequence in embedding space. Given a generated code sequence $C = (c_1, \dots, c_n)$ and a reference code sequence $R = (r_1, \dots, r_m)$, let $\mathbf{e}(c_i)$ and $\mathbf{e}(r_j)$ denote their contextual embeddings. The similarity between tokens is computed using cosine similarity:

\[
\text{sim}(c_i, r_j) = \frac{\mathbf{e}(c_i) \cdot \mathbf{e}(r_j)}{\|\mathbf{e}(c_i)\| \|\mathbf{e}(r_j)\|}
\]

We report Precision, Recall, F1, and F3, emphasizing F3 for its recall-focused evaluation of functional completeness.

\textit{Levenshtein Distance:} Levenshtein distance~\cite{levenshtein1966binary, 11029951} measures the minimum number of character-level edits (insertions, deletions, substitutions) required to transform generated code into the reference implementation. It provides a surface-level measure of textual similarity, allowing us to quantify how closely the generated code matches the reference in terms of syntax and structure, independent of
semantic correctness.

\subsubsection{Implementation} 
\name is implemented as a standalone tool targeting Python. To ensure safe and reproducible execution, all generated code was executed in isolated \texttt{Docker} containers (version~28.1.1). A pilot study on 20 randomly selected programs was used to determine the number of refinement loops. Consistent with prior work~\cite{le-cong-etal-2025-llms,ruan2025specrover}, performance improved primarily within the first three iterations, with only sporadic gains beyond that. Accordingly, we cap both execution-driven and review-driven refinement loops at three iterations per sub-task, balancing generation quality and computational cost. Open-source LLMs were run locally using \texttt{Ollama}, and closed-source models were accessed via official APIs. All experiments ran on a workstation with a 28-core Intel\textsuperscript{\textregistered}~Xeon\textsuperscript{\textregistered}~w7-3465X CPU (2.50~GHz), 256~GB RAM, and an NVIDIA~RTX~6000~Ada GPU under Ubuntu~24.04.2~LTS.

\subsection{Results}
\subsubsection{Ablation Study}
\label{sec:ablation}
\begin{table*}[!htpb]
\centering
\small
\caption{Performance comparison of Gemma3 variants on \namebenchissues dataset.}
\label{tab:ablation}
\vspace{-0.5em}

\begin{tabularx}{\textwidth}{p{0.17\textwidth} *{7}{R} *{7}{R}}
\toprule
\multirow{2}{*}{\textbf{Model}} 
& \multicolumn{7}{c}{\textbf{Buggy Code}} 
& \multicolumn{7}{c}{\textbf{Patched Code}} \\
\cmidrule(lr){2-8}
\cmidrule(lr){9-15}

& \textbf{ER} & \textbf{SR} & \textbf{P} & \textbf{R} & \textbf{F1} & \textbf{F3} & \textbf{LD}
& \textbf{ER} & \textbf{SR} & \textbf{P} & \textbf{R} & \textbf{F1} & \textbf{F3} & \textbf{LD} \\

\midrule

$\name_{\text{w/o refinement}}$
& 0.68 & 0.66 & 0.96 & 0.99 & 0.98 & 0.99 & 81.09
& 0.56 & 0.42 & 0.96 & 0.99 & 0.98 & 0.99 & 85.25 \\

$\name_{\text{w/o guidance}}$
& 0.70 & 0.68 & 0.95 & 0.97 & 0.96 & 0.97 & 70.64
& 0.68 & 0.46 & 0.92 & 0.94 & 0.93 & 0.94 & 120.29 \\

$\name_{\text{with guidance}}$
& 0.96 & 0.62 & 0.96 & 0.99 & 0.97 & 0.99 & 81.19
& 0.56 & 0.46 & 0.91 & 0.93 & 0.92 & 0.93 & 134.10 \\

\bottomrule
\end{tabularx}

\vspace{0.3em}
\begin{flushleft}
\footnotesize
\textbf{Note:} ER = Execution Rate; SR = Success Rate; 
P = Precision; R = Recall; 
F1/F3 = CodeBERTScore variants; LD = Levenshtein Distance (lower is better).
\end{flushleft}
\end{table*}
Table~\ref{tab:ablation} presents an ablation of guidance and refinement on the \namebenchissues dataset, with all variants using the same base prompt.

For buggy code, guidance improves executability, increasing ER from 0.70 to 0.96, indicating that guidance helps generate runnable programs by providing structural context. The slightly lower SR (0.62 vs. 0.68) suggests that improving executability does not always preserve semantic correctness, especially for open-source models with constrained context windows. For patched code, the trends are nuanced. While ER is highest without guidance, SR remains comparable across settings. This is expected, as buggy code in GitHub Issues is often explicitly available, making executable buggy-code reproduction easier than patch generation, which requires deeper reasoning beyond surface-level cues. CodeBERTScore for buggy code remains consistent across variants, suggesting that guidance preserves structural similarity. Similarity metrics are computed on successfully generated cases. For patched code, under this conditioning, guided variants exhibit lower F3 and higher Levenshtein distance, indicating more diverse but functionally valid patches.

Overall, guidance significantly improves executability, while patched-code correctness remains challenging due to the limitations of open-source models and the inherent difficulty of patch synthesis.

\subsubsection{RQ1 (Effectiveness of LLMs  for Code Replication)}
\label{subsubsec:RQ1}
\begin{figure*}[!t]
    \centering

    \begin{minipage}[t]{0.49\textwidth}
        \centering
        \includegraphics[width=\linewidth]{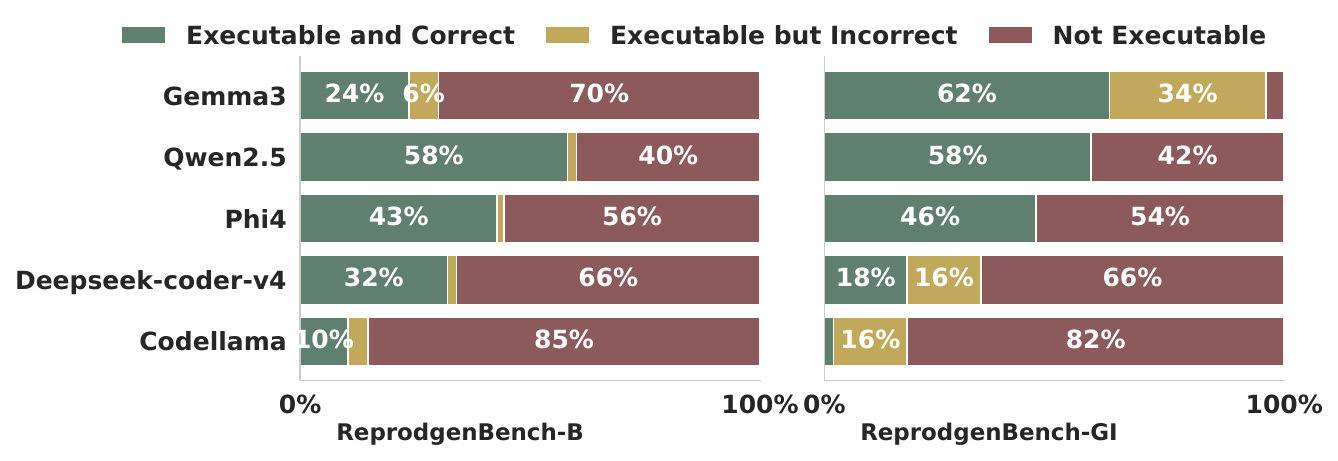}
        \subcaption{Buggy code}
        \label{fig:buggy-er-sr}
    \end{minipage}
    \hfill
    \begin{minipage}[t]{0.49\textwidth}
        \centering
        \includegraphics[width=\linewidth]{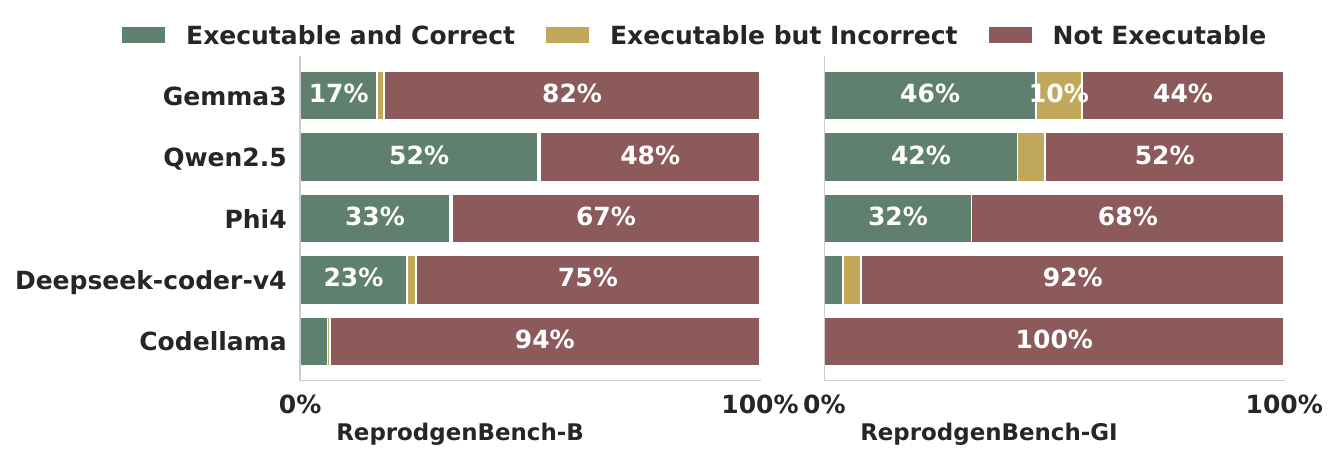}
        \subcaption{Patched code}
        \label{fig:patched-two-panel-b}
    \end{minipage}

    \caption{Execution and success ratios for buggy and patched code generation on \textit{\namebenchbase} and \textit{\namebenchissues}. Each bar shows the proportion of outputs that are executable and correct, executable but incorrect, and not executable.}
    \label{fig:patched-two-panel-overall}
\end{figure*}

Table~\ref{tab:RQ1} and Figure~\ref{fig:patched-two-panel-overall} report the effectiveness of \name on \namebenchbase and \namebenchissues using Execution Rate (ER), Success Rate (SR), CodeBERTScore, and Levenshtein Distance (LD). For all open-source models, we use Qwen~3 as the judge LLM.

\begin{table*}[htpb]
\centering
\footnotesize
\setlength{\tabcolsep}{3pt}
\renewcommand{\arraystretch}{0.95}
\caption{Comparison of model performance on buggy and patched code across datasets.}
\label{tab:RQ1}
\vspace{-0.5em}

\resizebox{\textwidth}{!}{%
\begin{tabular}{p{0.14\textwidth} p{0.15\textwidth} *{10}{r}}
\toprule
\multirow{2}{*}{\textbf{Dataset}} 
& \multirow{2}{*}{\textbf{Model}}
& \multicolumn{5}{c}{\textbf{Buggy Code}}
& \multicolumn{5}{c}{\textbf{Patched Code}} \\
\cmidrule(lr){3-7}
\cmidrule(lr){8-12}
& & \textbf{P} & \textbf{R} & \textbf{F1} & \textbf{F3} & \textbf{LD}
  & \textbf{P} & \textbf{R} & \textbf{F1} & \textbf{F3} & \textbf{LD} \\
\midrule

\multirow{5}{*}{\makecell[l]{\namebenchbase}}
& Gemma 3
& 0.90 & 0.80 & 0.84 & 0.80 & 756.10
& \cellcolor{gray!15}0.90 & 0.78 & 0.83 & 0.79 & 849.00 \\

& Qwen2.5
& \cellcolor{gray!15}0.91 & \cellcolor{gray!15}0.82 & \cellcolor{gray!15}0.86 & \cellcolor{gray!15}0.83 & \cellcolor{gray!15}715.80
& \cellcolor{gray!15}0.90 & \cellcolor{gray!15}0.80 & \cellcolor{gray!15}0.84 & \cellcolor{gray!15}0.80 & 847.10 \\

& Phi4
& 0.82 & 0.81 & 0.81 & 0.81 & 861.50
& 0.86 & 0.79 & 0.82 & 0.79 & 1058.10 \\

& DeepSeek-Coder-v4
& 0.89 & 0.79 & 0.84 & 0.80 & 831.70
& 0.89 & 0.75 & 0.81 & 0.76 & 952.80 \\

& Code Llama
& 0.86 & 0.74 & 0.79 & 0.75 & 870.01
& 0.89 & 0.74 & 0.81 & 0.75 & \cellcolor{gray!15}759.84 \\

\midrule

\multirow{5}{*}{\makecell[l]{\namebenchissues}}
& Gemma 3
& \cellcolor{gray!15}0.96 & \cellcolor{gray!15}0.99 & \cellcolor{gray!15}0.97 & \cellcolor{gray!15}0.99 & \cellcolor{gray!15}81.19
& 0.91 & 0.93 & 0.92 & 0.93 & 134.10 \\

& Qwen2.5
& 0.93 & 0.96 & 0.95 & 0.96 & 127.82
& 0.90 & 0.95 & 0.92 & 0.94 & 184.36 \\

& Phi4
& 0.72 & 0.76 & 0.74 & 0.76 & 206.20
& 0.77 & 0.82 & 0.79 & 0.81 & 241.65 \\

& DeepSeek-Coder-v4
& 0.54 & 0.55 & 0.55 & 0.55 & 243.37
& \cellcolor{gray!15}0.97 & \cellcolor{gray!15}0.98 & \cellcolor{gray!15}0.97 & \cellcolor{gray!15}0.98 & \cellcolor{gray!15}61.50 \\

& Code Llama
& 0.61 & 0.61 & 0.61 & 0.61 & 182.88
& -- & -- & -- & -- & -- \\

\bottomrule
\end{tabular}%
}

\begin{flushleft}
\footnotesize
\textbf{Note:} Precision, Recall, F1, and F3 are computed using CodeBERTScore. LD = Levenshtein Distance (lower is better). Gray cells indicate the best value in each metric column.
\vspace{-2em}
\end{flushleft}
\end{table*}

Across open-source models, Qwen~2.5 and Gemma~3 show the strongest overall performance. On \namebenchbase, Qwen~2.5 achieves the highest buggy-code SR, with 122 out of 203 generated programs judged executable and behaviorally consistent with the target. Gemma~3 and Phi-4 also achieve competitive executability. On \namebenchissues, Gemma~3 achieves the highest buggy-code ER, with 48 out of 50 replicated programs executable, followed by Qwen~2.5. These results indicate that execution reliability varies substantially across open-source models and does not scale monotonically with model size. This difference is partly explained by the nature of the datasets: GitHub Issues typically follow structured templates for reporting bugs and expected behavior, providing clearer context for LLMs to replicate, whereas Stack Overflow posts are more unstructured and discussion-driven, making replication more challenging.

The results also show a consistent gap between buggy-code replication and patched-code replication. Buggy code is often directly available or strongly implied in Stack Overflow posts and GitHub Issues, making it easier for models to reconstruct executable variants. In contrast, patched-code replication requires inferring the intended fix and producing code that matches the expected post-fix behavior. In addition, \name follows a sequential guidance pipeline: if an intermediate artifact generated during the refinement or review loop is incorrect or fails validation, it is discarded and not propagated to the next phase. While this prevents error accumulation, it also reduces the number of valid artifacts available for patched-code replication. These factors explain why patched-code ER and SR are generally lower than buggy-code ER and SR across models.

Beyond executability and behavioral success, CodeBERTScore captures semantic alignment between replicated and reference code, while LD measures structural divergence. We compute these metrics only over executable outputs, ensuring that similarity comparisons are meaningful. Models with stronger ER and SR generally also show stronger CodeBERTScore and lower LD. For example, Qwen~2.5 achieves the strongest buggy-code SR on \namebenchbase and also obtains the lowest buggy-code LD among open-source models. However, high similarity alone does not guarantee behavioral success, especially for patched code, reinforcing the need to report ER, SR, CodeBERTScore, and LD together.

\finding{
Among open-source models, Qwen~2.5 and Gemma~3 achieve the strongest overall performance under the \name pipeline, while Code Llama and DeepSeek-Coder exhibit the weakest execution and success rates. Buggy-code replication is consistently more reliable than patched-code replication across all models.
}

\subsubsection{RQ2 (Effectiveness of Reviewer LLM)}
\label{subsubsec:RQ2}

Table~\ref{tab:RQ2} shows the reliability of reviewer LLMs across multiple sub-tasks derived from \namebenchbase, including CI, FR, SCoT, and generated code. Performance is measured using Precision, Recall, and F1-score against human-annotated ground truth, together with Coverage, which captures the proportion of instances with valid, parsable outputs. To ensure fair comparison, all metrics are computed only on valid outputs and reported as coverage-weighted F1-scores. This selection is critical, as SR directly depends on the reliability of the reviewer LLM.
\begin{table*}[!htbp]
\centering
\small
\caption{Evaluation of Reviewer LLMs on buggy and patched code reviews across multiple sub-tasks.}
\label{tab:RQ2}
\vspace{-0.5em}

\begin{tabularx}{\textwidth}{
  p{0.10\textwidth}
  *{8}{R}
  @{\hspace{1em}}
  *{8}{R}
  @{\hspace{1em}}
  *{2}{R}
}
\toprule
\multirow{3}{*}{\textbf{Model}} &
\multicolumn{8}{c}{\textbf{Buggy Review}} &
\multicolumn{8}{c}{\textbf{Patched Review}} &
\multicolumn{2}{c}{} \\
\cmidrule(lr){2-9}
\cmidrule(lr){10-17}

& \multicolumn{2}{c}{\textbf{CI}} &
  \multicolumn{2}{c}{\textbf{FR}} &
  \multicolumn{2}{c}{\textbf{SCoT}} &
  \multicolumn{2}{c}{\textbf{Code}} &
  \multicolumn{2}{c}{\textbf{CI}} &
  \multicolumn{2}{c}{\textbf{FR}} &
  \multicolumn{2}{c}{\textbf{SCoT}} &
  \multicolumn{2}{c}{\textbf{Code}} &
  \multicolumn{2}{c}{\textbf{Average}} \\

\cmidrule(lr){2-3}
\cmidrule(lr){4-5}
\cmidrule(lr){6-7}
\cmidrule(lr){8-9}
\cmidrule(lr){10-11}
\cmidrule(lr){12-13}
\cmidrule(lr){14-15}
\cmidrule(lr){16-17}
\cmidrule(lr){18-19}

& \textbf{Cov.} & \textbf{F1} &
  \textbf{Cov.} & \textbf{F1} &
  \textbf{Cov.} & \textbf{F1} &
  \textbf{Cov.} & \textbf{F1} &
  \textbf{Cov.} & \textbf{F1} &
  \textbf{Cov.} & \textbf{F1} &
  \textbf{Cov.} & \textbf{F1} &
  \textbf{Cov.} & \textbf{F1} &
  \textbf{Cov.} & \textbf{F1} \\
\midrule

% GPT-4o
% & 1.00 & 0.98 & 1.00 & 0.99 & 1.00 & 0.92 & 1.00 & 0.79
% & 1.00 & 1.00 & 1.00 & 0.99 & 1.00 & 0.99 & 1.00 & \cellcolor{gray!60}0.78
% & 1.00 & \cellcolor{gray!60}0.93 \\

% \arrayrulecolor{gray!30}\midrule\arrayrulecolor{black}

Gemma~3
& 1.00 & 0.97 & 1.00 & 0.98 & 1.00 & 0.90 & 1.00 & 0.77
& 1.00 & 0.98 & 1.00 & 0.97 & 1.00 & 0.93 & 0.97 & 0.54
& 1.00 & 0.88 \\

Qwen~3
& 1.00 & 0.98 & 1.00 & 0.98 & 1.00 & 0.92 & 0.97 & 0.78
& 0.89 & 0.89 & 1.00 & 0.97 & 0.95 & 0.89 & 0.88 & 0.50
& 0.96 & 0.86 \\

Phi-4
& 0.97 & 0.93 & 0.97 & 0.94 & 0.99 & 0.89 & 0.99 & 0.70
& 0.99 & 0.97 & 0.99 & 0.96 & 0.99 & 0.91 & 0.98 & 0.44
& 0.99 & 0.84 \\

Llama~4
& 0.98 & 0.93 & 0.95 & 0.92 & 0.93 & 0.83 & \cellcolor{gray!15}0.62 & 0.47
& 0.74 & 0.73 & 0.74 & 0.71 & 0.93 & 0.85 & 0.88 & 0.72
& 0.85 & 0.77 \\

Code~Llama
& 1.00 & 0.50 & 1.00 & 0.50 & 1.00 & 0.78 & 0.78 & \cellcolor{gray!15}0.15
& 0.98 & 0.92 & 0.96 & 0.69 & 0.88 & 0.59 & 0.77 & \cellcolor{gray!15}0.42
& 0.92 & \cellcolor{gray!15}0.57 \\

\bottomrule
\end{tabularx}

\begin{flushleft}
\footnotesize
\textbf{Note:} Cov.~=~Coverage; CI~=~Code~Intent; FR~=~Functional~Requirements; 
SCoT~=~Structured chain of thought. 
\colorbox{gray!60}{\rule{0pt}{6pt}\rule{6pt}{0pt}}~best, 
\colorbox{gray!15}{\rule{0pt}{3pt}\rule{3pt}{0pt}}~worst values.
\end{flushleft}
\end{table*}

Across open-source reviewer LLMs, Gemma~3 and Qwen~3 achieve the strongest overall performance, with average coverage-weighted F1-scores of 0.88 and 0.86, respectively. Both models perform consistently well on text-centric subtasks (CI, FR, SCoT), achieving F1-scores above 0.90 in most settings, indicating reliable semantic interpretation of problem descriptions. In contrast, Phi-4 and Llama~4 exhibit moderate performance, while smaller models such as Code Llama (7B) perform poorly, particularly on code-level evaluation.

A consistent trend across all models is the performance gap between textual and code-level review. While CI, FR, and SCoT generally achieve high F1-scores, performance drops on the Code subtask, especially for patched code. For example, Gemma~3 declines from 0.77 F1 on buggy code to 0.54 on patched code, and Qwen~3 drops from 0.78 to 0.50. This reflects the increased difficulty of evaluating patched code, which requires reasoning over longer contexts and subtle behavioral changes rather than direct semantic matching.

Coverage further differentiates reviewer reliability. Models with higher Coverage generally achieve higher F1-scores, indicating that adherence to the expected labeling schema is strongly correlated with accuracy. Gemma~3 maintains near-perfect Coverage across all tasks, while Qwen~3 shows slightly lower average Coverage (0.96), primarily due to occasional formatting deviations. Code Llama (7B) exhibits both low Coverage and low F1, limiting its suitability as an automated reviewer.

Although Gemma~3 achieves the highest average F1, Qwen~3 achieves comparable reviewer performance while offering lower inference costs and faster response times. We therefore select Qwen~3 as the standardized reviewer LLM for computing SR across all generator models. This ensures that differences in SR reflect generator performance rather than variations in the reviewer-LLM.

\finding{
Gemma~3 and Qwen~3 are the most reliable open-source reviewer LLMs, achieving the highest coverage-weighted F1-scores. Reviewer performance is consistently strong on textual subtasks but degrades significantly for code-level evaluation, especially for patched code.}

\subsubsection{RQ3 (Efficiency)}
\label{subsubsec:RQ3}
\begin{table*}[!htpb]
\centering
\small
\caption{Efficiency analysis of buggy and patched code generation across open source models in \namebenchbase.}
\label{tab:RQ3-efficiency}
\vspace{-0.5em}

\begin{tabularx}{\textwidth}{
  p{0.16\textwidth}
  *{5}{R}
  @{\hspace{1em}}
  *{5}{R}
}
\toprule
\multirow{2}{*}{\textbf{Model}} &
\multicolumn{5}{c}{\textbf{Buggy Code}} &
\multicolumn{5}{c}{\textbf{Patched Code}} \\
\cmidrule(lr){2-6}
\cmidrule(lr){7-11}

& \textbf{Time} & \textbf{Input} & \textbf{Output} & \textbf{EI} & \textbf{RI}
& \textbf{Time} & \textbf{Input} & \textbf{Output} & \textbf{EI} & \textbf{RI} \\
\midrule

% Claude Sonnet~4.5
% & 18.25 & 4688.03 & 1284.87 & 1.60 & 2.17
% & 13.69 & 4233.10 & 1015.95 & 1.33 & 2.12 \\

% \arrayrulecolor{gray!30}\midrule\arrayrulecolor{black}

Qwen~2.5 Coder
& 20.21 & 4381.28 & 541.24 & 2.18 & 1.91
& 14.71 & 3594.77 & 351.97 & 1.20 & 2.06 \\

Gemma~3
& 17.37 & 4736.61 & 508.97 & 2.26 & 1.98
& 13.24 & 3434.47 & 350.98 & 1.38 & 1.95 \\

DeepSeek-Coder-V2
& 10.38 & 5146.31 & 552.80 & 2.49 & 1.83
& 6.33 & 4836.85 & 368.60 & 1.46 & 2.17 \\

Phi-4
& 18.84 & 5546.26 & 929.12 & 2.55 & 1.72
& 11.69 & 4360.34 & 498.40 & 1.35 & 2.08 \\

Code Llama
& 3.18 & 3412.73 & 298.96 & 2.28 & 1.19
& 5.96 & 7484.30 & 527.10 & 1.89 & 2.11 \\

\bottomrule
\end{tabularx}

\begin{flushleft}
\footnotesize
\textbf{Note:} Time~=~average generation time (seconds); Input/Output~=~token counts; EI~=~execution refinement iterations; RI~=~review refinement iterations.
\end{flushleft}
\vspace{-2em}
\end{table*}

We evaluate efficiency using average generation time, token usage, and refinement behavior, measured by execution iterations (EI) and review iterations (RI). All metrics are computed per Stack Overflow question by aggregating the full generation process until either an executable artifact is produced or the maximum number of refinement attempts is exhausted.

Among open-source models, Gemma~3 and Qwen~2.5 Coder provide the strongest efficiency--reliability balance. Gemma~3 is faster than Qwen~2.5 Coder for both buggy code (17.37\,sec vs. 20.21\,sec) and patched code (13.24\,sec vs. 14.71\,sec), while maintaining comparable refinement behavior. Qwen~2.5 Coder requires slightly longer generation time but achieves the lowest patched-code EI (1.20), indicating more stable convergence once the buggy artifact is available. In contrast, DeepSeek-Coder-V2 and Code Llama exhibit lower raw generation latency but require more execution refinements, reflecting less stable convergence. Phi-4 incurs relatively high token usage and refinement overhead, particularly for buggy-code generation. We further analyze refinement behavior on \namebenchissues. Refinement has limited impact for buggy-code replication, as the buggy code is often directly available, leading to high ER and SR even without refinement (e.g., ER=0.68, SR=0.66 for Gemma~3 without refinement). In contrast, refinement is more beneficial for patched code, where it improves SR (e.g., SR increases from 0.42 to 0.46), indicating that iterative feedback is more important for achieving behaviorally correct patches.

Overall, efficiency in \name is governed by both inference latency and convergence behavior. Faster models are not necessarily more efficient if they require additional refinement iterations or produce fewer usable artifacts.

\finding{
Gemma~3 and Qwen~2.5 Coder offer the best efficiency--reliability trade-off among open-source models: Gemma~3 is faster overall, while Qwen~2.5 Coder requires the fewest patched-code execution refinements.
}

\subsection{Lessons learned}
Based on our findings, we summarize key lessons for using LLMs as reviewers and code generators.

\circleB{1} \textbf{\name can generate patched code without accepted answers, but accepted answers improve alignment with intended fixes.}
On 29 posts from \namebenchissues, Gemma~3 achieved high executability (ER=0.90) without using accepted answers, but behavioral success was low (SR=0.083), yielding only 11 correct patched-code replications (F1=0.854, F3=0.861, and LD=119.14). This suggests that accepted answers are not required for patch generation, but they provide important grounding for producing fixes that match the developer's intent.

\circleB{2} \textbf{LLMs perform well on textual tasks but struggle with code-level analysis.}
LLMs achieve high performance on CI, FR, and SCoT extraction, but their performance drops for code-level evaluation, particularly for patched code (RQ2 Section~\ref{subsubsec:RQ2}). This indicates that textual understanding alone is insufficient for reliable code assessment. Practitioners should pair LLM reviewers with lightweight static or dynamic analyzers, while researchers should develop hybrid evaluation methods that combine semantic similarity with execution feedback.

\circleB{3} \textbf{Guided prompting improves executability and stabilizes replication.}
Guidance improves ER while maintaining comparable SR, suggesting that structured artifacts such as CI, FR, and SCoT help models produce runnable replications without degrading accepted answer alignment (RQ1 Section~\ref{subsubsec:RQ1}). Practitioners should design prompts that explicitly encode task intent and constraints, while researchers should combine structured prompting with execution-based validation and refinement.

\circleB{4} \textbf{Open-source LLMs show inconsistent schema adherence.}
Some open source models generate unsupported labels or formatting deviations even under deterministic settings, indicating that fixed temperature and seed do not guarantee schema-compliant outputs in automated pipelines. Practitioners should incorporate structured output, output validation, and constrained decoding, while researchers should investigate structured generation techniques to improve consistency and adherence.

\circleB{5} \textbf{Open-source LLMs struggle with dependency inference.}
LLMs often fail to infer correct library versions, dependency constraints, and environment configurations needed to reproduce Stack Overflow bugs, particularly for open-source models. This highlights the difficulty of reasoning about version-specific behavior from pretrained knowledge alone. Practitioners should integrate retrieval or web-based tools for dependency resolution, while researchers should explore retrieval- and tool-augmented approaches for improving environment inference.

\section{Related Work}
\label{sec:relatedWork}

Research on patch generation for data science code involves bug analysis, program repair, assessment methods, and the use of LLMs. This section reviews key advances in these areas.

\paragraph{\textbf{Empirical Study in Data Science (DS)}}
Comprehensive empirical studies have offered valuable insights into bug characteristics, mistake patterns, and challenges specific to data science code. Analyses of Jupyter notebooks show that data-centric bugs are common and often coupled with weak testing practices~\cite{SantanaTOSEM}. Ahmed et~al.~\cite{ahmed2023characterizing} compared Python and~R analytics programs, revealing diverse bug types and fix strategies across languages. Zhao and Fard examined LLM capabilities for~R code intelligence~\cite{zhao2024llmsR} and identified vulnerable code entities~\cite{zhao2024vulnerableR}, underscoring language-specific challenges in patch generation. Other work has focused on how errors emerge and are repaired in notebook environments. Wang et~al.~\cite{wang2025machine} found that dependency conflicts, data issues, and API misuse are major causes of notebook failures. Rawal et~al.~\cite{rawal2025hints} showed that hints influence how developers locate and fix bugs in Python and text-based representations. Chen et~al.~\cite{chen2025towards} provided a taxonomy of fine-grained programming errors and repair patterns in data science workflows, while Yang et~al.~\cite{yang2025towards} emphasized the role of non-functional faults, particularly performance issues, in widely used data science libraries.

\paragraph{\textbf{Evaluating LLMs on Coding Tasks}}
Patch generation for data science code has progressed more slowly than for general-purpose languages due to dynamic semantics and library-specific APIs. Knowledge-enhanced methods such as DSrepair~\cite{ouyang2025knowledge}, which combine API usage graphs with AST-based bug detection, are improving this process. Benchmarks like DS-1000 show that domain-specific approaches outperform general LLM-based repairs. Controlled studies using models such as CodeLlama, StarCoder, and DeepSeek-Coder indicate that prompt design and model choice strongly affect patch quality~\cite{nascimento2025effective}. Ruan et~al.'s SpecRover~\cite{ruan2025specrover} uses LLMs for code intent extraction, while frameworks like AutoCodeRover~\cite{AutoCodeRover2024} explore autonomous, iterative repair. Recent work, such as RepGen~\cite{shah2025imitation}, further advances this direction by introducing an agentic generate--validate--refine workflow for reproducing deep learning bugs using structured bug reports and repository-level context. Complementary to repair and reproduction, ZS4C~\cite{kabir2025zs4c} and SelfPiCo~\cite{xue2024selfpico} address executability challenges in incomplete code snippets from Q\&A sites: ZS4C synthesizes compilable code by inferring missing imports and iteratively resolving compilation errors, while SelfPiCo enables partial-code execution by inferring missing runtime values, definitions, and dependencies from execution feedback. Du et~al.~\cite{du2024evaluating} assess class-level code generation under adversarial settings, and Zhao and Fard~\cite{zhao2024llmsR} highlight the need for repair models tailored to less-dominant languages. Broader reviews trace the shift from search- and constraint-based repair to learning-driven approaches~\cite{fei2025patch} and emphasize challenges in vulnerability repair and semantic patch evaluation, particularly for R~\cite{zhao2024llmsR,zhao2024vulnerableR}. Overall, these studies demonstrate growing progress in LLM-based repair, reproduction, and executability support, while our work focuses on reconstructing executable buggy--patched data science programs from informal developer discussions with both execution-based validation and semantic faithfulness assessment.

\paragraph{\textbf{Patch Correctness Assessment}}
The reliability of automated patch generation depends on robust, scalable correctness assessment. Traditional strategies include developer/labeller annotation, automated test suites, and static/dynamic analysis~\cite{le2019reliability, PatchCorrectness2018}. However, empirical reviews show no method is sufficient across all cases and domains~\cite{fei2025patch}. Randomized test augmentation~\cite{molina2024improving} and large-scale automated surveys~\cite{PatchCorrectness2018, fei2025patch} highlight ongoing challenges with false positives/negatives. Recent work proposes hybrid correctness assessment: PatchZero explores zero-shot LLM-based approaches to judge patch correctness using natural language descriptions and improved context understanding~\cite{zhou2023patchzero}, while Molina et al. combine random testing with LLM judgment for higher precision on real-world datasets~\cite{molina2024improving}. Ouyang et al. demonstrate that domain-specific knowledge and verification yield more accurate and relevant patch assessments for DS code~\cite{ouyang2025knowledge}. Such multi-faceted approaches are particularly vital in DS code, where test coverage is typically weak and core logic is data-dependent.

\paragraph{\textbf{ Evaluating LLMs as Judge}}
LLM-as-a-Judge frameworks have introduced new possibilities for automated patch validation, offering ways to complement or even reduce human evaluation effort. Wang et~al.~\cite{WangJudgeLLM} compared LLM-based judgment with human assessment and found that while LLMs show strong potential, they are not yet fully reliable for complex, real-world patch evaluation. This line of work has encouraged wider use of LLM-assisted review in empirical repair pipelines, as summarized by Fei et~al.~\cite{fei2025patch}.

Earlier research has examined bug characterization, LLM-based code generation, and patch correctness assessment separately, but no prior framework integrates these components into a single, systematic workflow. Our approach, \name fills this gap by combining structured intermediate representations with a collaborative refinement process driven by LLMs. Combines automated validation with expert human review, creating a reliable method for generating and verifying executable buggy–patched code from informal developer Q\&A forums. This design makes \name a strong basis for future research and tool development in debugging and automated repair for data science programs.
\section{Threats to Validity}
\label{sec:threats}

\textbf{Internal threats.} LLMs can exhibit stochastic behavior, where small changes in decoding may lead to different outputs. To control variability, we fixed all sampling parameters across runs and assessed stability by rerunning a random subset of instances, observing no substantial variation in aggregate metrics. Labeling bias was reduced through independent annotation by experts with disagreement resolution~\cite{Islam2019FSE, biswas2022}, and all models were evaluated using identical prompts. Because the evaluation benchmarks are derived from Stack Overflow posts and GitHub Issues, partial overlap with training data is possible; we mitigated this by selecting diverse posts across multiple libraries that emphasize mock data and dependency inference rather than verbatim code reuse. A further threat arises from the semantic filtering stage, which excludes many posts that do not clearly describe reproducible buggy behavior. This filtering may omit some valid but ambiguous bugs, potentially limiting coverage; however, our goal is not exhaustiveness, but the construction of a high-confidence benchmark of executable buggy-patched pairs, and all retained instances were subsequently validated by human experts, reducing the likelihood of filter-induced bias.

\textbf{External threats:} The evaluation benchmarks in our study cover Stack Overflow posts and GitHub Issues related to data science libraries, following trends~\cite{soTagtrends}. While representative of real-world data science issues, it may not generalize to other domains or languages. To reduce this bias, we included multiple bug categories, logical, functional, and performance-related, spanning seven major libraries, and the results should be interpreted within this scope.

\section{Conclusion and Future Work}
\label{sec:conclusion}
We presented \name, a collaborative framework that leverages LLMs to generate executable buggy and patched data science-related programs from informal Stack Overflow and GitHub Issue discussions. \name extracts structured semantic guidance, including CI, FR, and SCoT, and iteratively refines generated artifacts through an LLM-based review loop to resolve ambiguity and missing context. In our experimental evaluation, \name consistently produced executable buggy-patched code pairs with high execution success rates and strong semantic correctness. We additionally release a benchmark and a community-driven leaderboard to support reproducible evaluation and transparent comparison across models. Together, these contributions establish \name as a reproducible and extensible foundation for studying LLM-based bug reproduction and repair in real-world data science programs. Future work includes extending \name to expand into additional programming domains, enabling retrieval-augmented generation (RAG) for API-aware synthesis, and exploring richer forms of collaborative code refinement.

\section{Data Availability}
\label{sec:datapackage}
The reproducibility package~\cite{ReprodGenRepository}, prompts~\cite{ReprodGenPrompts}, benchmarks~\cite{ReprodGenBench}, and leaderboard~\cite{ReprodGenGenLeaderboard} are available in an anonymous repository~\cite{ReprodGenRepository}, and we hope it serves as a useful resource for future research in this area.

\bibliographystyle{IEEEtran}
\bibliography{biblography}

@article{kabir2025zs4c,
  title={Zs4c: Zero-shot synthesis of compilable code for incomplete code snippets using llms},
  author={Kabir, Azmain and Wang, Shaowei and Tian, Yuan and Chen, Tse-Hsun and Asaduzzaman, Muhammad and Zhang, Wenbin},
  journal={ACM Transactions on Software Engineering and Methodology},
  volume={34},
  number={4},
  pages={1--30},
  year={2025},
  publisher={ACM New York, NY}
}

@inproceedings{xue2024selfpico,
  title={Selfpico: Self-guided partial code execution with llms},
  author={Xue, Zhipeng and Gao, Zhipeng and Wang, Shaohua and Hu, Xing and Xia, Xin and Li, Shanping},
  booktitle={Proceedings of the 33rd ACM SIGSOFT International Symposium on Software Testing and Analysis},
  pages={1389--1401},
  year={2024}
}

@article{shah2025imitation,
  title={Imitation Game: Reproducing Deep Learning Bugs Leveraging an Intelligent Agent},
  author={Shah, Mehil B and Rahman, Mohammad Masudur and Khomh, Foutse},
  journal={arXiv preprint arXiv:2512.14990},
  year={2025}
}

@article{yuan2024evaluating,
  title={Evaluating and improving chatgpt for unit test generation},
  author={Yuan, Zhiqiang and Liu, Mingwei and Ding, Shiji and Wang, Kaixin and Chen, Yixuan and Peng, Xin and Lou, Yiling},
  journal={Proceedings of the ACM on Software Engineering},
  volume={1},
  number={FSE},
  pages={1703--1726},
  year={2024},
  publisher={ACM New York, NY, USA}
}

@article{chao2024jailbreakbench,
  title={Jailbreakbench: An open robustness benchmark for jailbreaking large language models},
  author={Chao, Patrick and Debenedetti, Edoardo and Robey, Alexander and Andriushchenko, Maksym and Croce, Francesco and Sehwag, Vikash and Dobriban, Edgar and Flammarion, Nicolas and Pappas, George J and Tramer, Florian and others},
  journal={Advances in Neural Information Processing Systems},
  volume={37},
  pages={55005--55029},
  year={2024}
}

@misc{bigcode-leaderboard,
  title        = {BigCode Models Leaderboard},
  author       = {{BigCode}},
  year         = {2026},
  publisher    = {Hugging Face},
  howpublished = {\url{https://huggingface.co/spaces/bigcode/bigcode-models-leaderboard}},
  note         = {Accessed: 2026-04-20}
}

@article{zhang2025llm,
  title={Llm hallucinations in practical code generation: Phenomena, mechanism, and mitigation},
  author={Zhang, Ziyao and Wang, Chong and Wang, Yanlin and Shi, Ensheng and Ma, Yuchi and Zhong, Wanjun and Chen, Jiachi and Mao, Mingzhi and Zheng, Zibin},
  journal={Proceedings of the ACM on Software Engineering},
  volume={2},
  number={ISSTA},
  pages={481--503},
  year={2025},
  publisher={ACM New York, NY, USA}
}

@misc{soTagtrends,
  title        = {{Tag Trends Data Science Libraries}},
  howpublished = {\url{https://trends.stackoverflow.co/?tags=pandas\%2Cscipy\%2Cscikit-learn\%2Cnumpy\%2Cmatplotlib\%2Cdataframe\%2Cpyqt}},
  year         = {2025},
  note         = {[Online; accessed Apr-2026]}
}

@misc{ReprodGenPrompts,
	title = {{ReprodGen Prompts}},
	howpublished = "\url{https://github.com/reprodgen/reprodgen/tree/main/src/reprodbench/llm/prompts}",
	year = {2026}, 
	note = "[Online; accessed Apr-2026]"
}

@misc{ReprodGenRepository,
	title = {{ReprodGen Repository}},
	howpublished = "\url{https://github.com/reprodgen/reprodgen}",
	year = {2026}, 
	note = "[Online; accessed Apr-2026]"
}

@misc{ReprodGenBench,
	title = {{ReprodGen Bench}},
	howpublished = "\url{https://github.com/reprodgen/reprodgen/tree/main/data}",
	year = {2026}, 
	note = "[Online; accessed Apr-2026]"
}

@misc{ReprodGenGenLeaderboard,
	title = {{ReprodGen Leaderboard}},
	howpublished = "\url{https://github.com/reprodgen/reprodgen-leaderboard}",
	year = {2026}, 
	note = "[Online; accessed Apr-2026]"
}

@misc{ReprodgenSota,
	title = {{Output logs from AutoCodeRover and ArchCode }},
	howpublished = "\url{https://github.com/reprodgen/reprodgen/tree/main/sota-comparison}",
	year = {2026}, 
	note = "[Online; accessed Apr-2026]"
}

@misc{ReprodgenResults,
	title = {{Intermediate artifacts}},
	howpublished = "\url{https://github.com/reprodgen/reprodgen/tree/main/results/figures}",
	year = {2026}, 
	note = "[Online; accessed Apr-2026]"
}

@article{10.1145/3643763,
author = {Misu, Md Rakib Hossain and Lopes, Cristina V. and Ma, Iris and Noble, James},
title = {Towards AI-Assisted Synthesis of Verified Dafny Methods},
year = {2024},
issue_date = {July 2024},
publisher = {Association for Computing Machinery},
address = {New York, NY, USA},
volume = {1},
number = {FSE},
url = {https://doi.org/10.1145/3643763},
doi = {10.1145/3643763},
journal = {Proc. ACM Softw. Eng.},
month = jul,
articleno = {37},
numpages = {24}
}

@article{team2025gemma,
  title={Gemma 3 technical report},
  author={Team, Gemma and Kamath, Aishwarya and Ferret, Johan and Pathak, Shreya and Vieillard, Nino and Merhej, Ramona and Perrin, Sarah and Matejovicova, Tatiana and Ram{\'e}, Alexandre and Rivi{\`e}re, Morgane and others},
  journal={arXiv preprint arXiv:2503.19786},
  year={2025}
}

@article{zhu2024deepseek,
  title={Deepseek-coder-v2: Breaking the barrier of closed-source models in code intelligence},
  author={Zhu, Qihao and Guo, Daya and Shao, Zhihong and Yang, Dejian and Wang, Peiyi and Xu, Runxin and Wu, Y and Li, Yukun and Gao, Huazuo and Ma, Shirong and others},
  journal={arXiv preprint arXiv:2406.11931},
  year={2024}
}

@inproceedings{zhao2024llmsR,
  author = {Zixiao Zhao and Fatemeh Fard},
  title = {Do Current Language Models Support Code Intelligence for R Programming Language?},
  booktitle = {ACM Transactions on Software Engineering and Methodology},
  year = {2024}
}

@inproceedings{zhao2024vulnerableR,
  author = {Zixiao Zhao and Millon Madhur Das and Fatemeh Fard},
  title = {Studying Vulnerable Code Entities in R},
  booktitle = {Proceedings of the 32nd IEEE/ACM International Conference on Program Comprehension (ICPC)},
  pages = {328--332},
  year = {2024}
}

@inproceedings{ruan2025specrover,
  title={SpecRover: Code Intent Extraction via LLMs},
  author={Ruan, Haifeng and Zhang, Yuntong and Roychoudhury, Abhik},
  booktitle={2025 IEEE/ACM 47th International Conference on Software Engineering (ICSE)},
  pages={963--974},
  year={2025},
  organization={IEEE}
}

@inproceedings{han2024archcode,
  title={ArchCode: Incorporating Software Requirements in Code Generation with Large Language Models},
  author={Han, Hojae and Kim, Jaejin and Yoo, Jaeseok and Lee, Youngwon and Hwang, Seung-won},
  booktitle={Proceedings of the 62nd Annual Meeting of the Association for Computational Linguistics (Volume 1: Long Papers)},
  pages={13520--13552},
  year={2024}
}

@article{li2025structured,
  title={Structured chain-of-thought prompting for code generation},
  author={Li, Jia and Li, Ge and Li, Yongmin and Jin, Zhi},
  journal={ACM Transactions on Software Engineering and Methodology},
  volume={34},
  number={2},
  pages={1--23},
  year={2025},
  publisher={ACM New York, NY}
}

@inproceedings{le2019reliability,
  title={On reliability of patch correctness assessment},
  author={Le, Xuan-Bach D and Bao, Lingfeng and Lo, David and Xia, Xin and Li, Shanping and Pasareanu, Corina},
  booktitle={2019 IEEE/ACM 41st International Conference on Software Engineering (ICSE)},
  pages={524--535},
  year={2019},
  organization={IEEE}
}

@inproceedings{zhou-etal-2023-codebertscore,
    title = "{C}ode{BERTS}core: Evaluating Code Generation with Pretrained Models of Code",
    author = "Zhou, Shuyan  and
      Alon, Uri  and
      Agarwal, Sumit  and
      Neubig, Graham",
    editor = "Bouamor, Houda  and
      Pino, Juan  and
      Bali, Kalika",
    booktitle = "Proceedings of the 2023 Conference on Empirical Methods in Natural Language Processing",
    month = dec,
    year = "2023",
    address = "Singapore",
    publisher = "Association for Computational Linguistics",
    url = "https://aclanthology.org/2023.emnlp-main.859/",
    doi = "10.18653/v1/2023.emnlp-main.859",
    pages = "13921--13937"
}

@INPROCEEDINGS {11029951,
author = { Wang, Zhijie and Zhou, Zijie and Song, Da and Huang, Yuheng and Chen, Shengmai and Ma, Lei and Zhang, Tianyi },
booktitle = { 2025 IEEE/ACM 47th International Conference on Software Engineering (ICSE) },
title = {{ Towards Understanding the Characteristics of Code Generation Errors Made by Large Language Models }},
year = {2025},
volume = {},
ISSN = {},
pages = {2587-2599},
doi = {10.1109/ICSE55347.2025.00180},
url = {https://doi.ieeecomputersociety.org/10.1109/ICSE55347.2025.00180},
publisher = {IEEE Computer Society},
address = {Los Alamitos, CA, USA},
month =May}

@inproceedings{nascimento2025effective,
  title={How Effective are LLMs for Data Science Coding? A Controlled Experiment},
  author={Nascimento, Nathalia and Guimaraes, Everton and Chintakunta, Sai Sanjna and Boominathan, Santhosh Anitha},
  booktitle={2025 IEEE/ACM 22nd International Conference on Mining Software Repositories (MSR)},
  pages={211--222},
  year={2025},
  organization={IEEE}
}

@inproceedings{du2024evaluating,
  title={Evaluating large language models in class-level code generation},
  author={Du, Xueying and Liu, Mingwei and Wang, Kaixin and Wang, Hanlin and Liu, Junwei and Chen, Yixuan and Feng, Jiayi and Sha, Chaofeng and Peng, Xin and Lou, Yiling},
  booktitle={Proceedings of the IEEE/ACM 46th International Conference on Software Engineering},
  pages={1--13},
  year={2024}
}

@article{ouyang2025knowledge,
  title={Knowledge-Enhanced Program Repair for Data Science Code},
  author={Ouyang, Shuyin and Zhang, Jie M and Sun, Zeyu and Penuela, Albert Merono},
  journal={arXiv preprint arXiv:2502.09771},
  year={2025}
}

@article{WangJudgeLLM,
author = {Wang, Ruiqi and Guo, Jiyu and Gao, Cuiyun and Fan, Guodong and Chong, Chun Yong and Xia, Xin},
title = {Can LLMs Replace Human Evaluators? An Empirical Study of LLM-as-a-Judge in Software Engineering},
year = {2025},
issue_date = {July 2025},
publisher = {Association for Computing Machinery},
address = {New York, NY, USA},
volume = {2},
number = {ISSTA},
url = {https://doi.org/10.1145/3728963},
doi = {10.1145/3728963},
journal = {Proc. ACM Softw. Eng.},
month = jun,
articleno = {ISSTA086},
numpages = {23}
}

@article{SantanaTOSEM,
author = {De Santana, Taijara Loiola and Neto, Paulo Anselmo Da Mota Silveira and De Almeida, Eduardo Santana and Ahmed, Iftekhar},
title = {Bug Analysis in Jupyter Notebook Projects: An Empirical Study},
year = {2024},
issue_date = {May 2024},
publisher = {Association for Computing Machinery},
address = {New York, NY, USA},
volume = {33},
number = {4},
issn = {1049-331X},
url = {https://doi.org/10.1145/3641539},
doi = {10.1145/3641539},
journal = {ACM Trans. Softw. Eng. Methodol.},
month = apr,
articleno = {101},
numpages = {34}
}

@inproceedings{rawal2025hints,
  title={Hints Help Finding and Fixing Bugs Differently in Python and Text-based Program Representations},
  author={Rawal, Ruchit and P{\u{a}}durean, Victor-Alexandru and Apel, Sven and Singla, Adish and Toneva, Mariya},
  booktitle={2025 IEEE/ACM 47th International Conference on Software Engineering (ICSE)},
  pages={729--729},
  year={2025},
  organization={IEEE Computer Society}
}

@article{ahmed2023characterizing,
  title={Characterizing bugs in python and r data analytics programs},
  author={Ahmed, Shibbir and Wardat, Mohammad and Bagheri, Hamid and Cruz, Breno Dantas and Rajan, Hridesh},
  journal={arXiv preprint arXiv:2306.08632},
  year={2023}
}

@inproceedings{biswas2022,
author = {Biswas, Sumon and Wardat, Mohammad and Rajan, Hridesh},
title = {The art and practice of data science pipelines: A comprehensive study of data science pipelines in theory, in-the-small, and in-the-large},
year = {2022},
isbn = {9781450392211},
publisher = {Association for Computing Machinery},
address = {New York, NY, USA},
url = {https://doi.org/10.1145/3510003.3510057},
doi = {10.1145/3510003.3510057},
booktitle = {Proceedings of the 44th International Conference on Software Engineering},
pages = {2091–2103},
numpages = {13},
location = {Pittsburgh, Pennsylvania},
series = {ICSE '22}
}

@article{yang2025towards,
  title={Towards Understanding Performance Bugs in Popular Data Science Libraries},
  author={Yang, Haowen and Li, Zhengda and Zhong, Zhiqing and Tang, Xiaoying and He, Pinjia},
  journal={Proceedings of the ACM on Software Engineering},
  volume={2},
  number={FSE},
  pages={2335--2358},
  year={2025},
  publisher={ACM New York, NY, USA}
}

@article{chen2025towards,
  title={Towards Understanding Fine-Grained Programming Mistakes and Fixing Patterns in Data Science},
  author={Chen, Wei-Hao and Cheoh, Jia Lin and Keim, Manthan and Brunswicker, Sabine and Zhang, Tianyi},
  journal={Proceedings of the ACM on Software Engineering},
  volume={2},
  number={FSE},
  pages={1824--1846},
  year={2025},
  publisher={ACM New York, NY, USA}
}

@article{wang2025machine,
  title={Why do Machine Learning Notebooks Crash? An Empirical Study on Public Python Jupyter Notebooks},
  author={Wang, Yiran and Meijer, Willem and Lopez, Jose Antonio Hernandez and Nilsson, Ulf and Varro, Daniel},
  journal={IEEE Transactions on Software Engineering},
  year={2025},
  publisher={IEEE}
}

@inproceedings{AutoCodeRover2024,
author = {Zhang, Yuntong and Ruan, Haifeng and Fan, Zhiyu and Roychoudhury, Abhik},
title = {AutoCodeRover: Autonomous Program Improvement},
year = {2024},
isbn = {9798400706127},
publisher = {Association for Computing Machinery},
address = {New York, NY, USA},
url = {https://doi.org/10.1145/3650212.3680384},
doi = {10.1145/3650212.3680384},
booktitle = {Proceedings of the 33rd ACM SIGSOFT International Symposium on Software Testing and Analysis},
pages = {1592–1604},
numpages = {13},
location = {Vienna, Austria},
series = {ISSTA 2024}
}

@inproceedings{PatchCorrectness2018,
author = {Xiong, Yingfei and Liu, Xinyuan and Zeng, Muhan and Zhang, Lu and Huang, Gang},
title = {Identifying patch correctness in test-based program repair},
year = {2018},
isbn = {9781450356381},
publisher = {Association for Computing Machinery},
address = {New York, NY, USA},
url = {https://doi.org/10.1145/3180155.3180182},
doi = {10.1145/3180155.3180182},
booktitle = {Proceedings of the 40th International Conference on Software Engineering},
pages = {789–799},
numpages = {11},
location = {Gothenburg, Sweden},
series = {ICSE '18}
}

@article{fei2025patch,
  title={Patch Correctness Assessment: A Survey},
  author={Fei, Zhiwei and Ge, Jidong and Li, Chuanyi and Wang, Tianqi and Li, Yuning and Zhang, Haodong and Huang, LiGuo and Luo, Bin},
  journal={ACM Transactions on Software Engineering and Methodology},
  volume={34},
  number={2},
  pages={1--50},
  year={2025},
  publisher={ACM New York, NY}
}

@article{zhou2023patchzero,
  title={PatchZero: Zero-shot automatic patch correctness assessment},
  author={Zhou, Xin and Xu, Bowen and Kim, Kisub and Han, DongGyun and Le-Cong, Thanh and He, Junda and Le, Bach and Lo, David},
  journal={arXiv preprint arXiv:2303.00202},
  year={2023}
}

@inproceedings{molina2024improving,
  title={Improving Patch Correctness Analysis via Random Testing and Large Language Models},
  author={Molina, Facundo and Copia, Juan Manuel and Gorla, Alessandra},
  booktitle={2024 IEEE Conference on Software Testing, Verification and Validation (ICST)},
  pages={317--328},
  year={2024},
  organization={IEEE}
}

@inproceedings{Islam2019FSE,
author = {Islam, Md Johirul and Nguyen, Giang and Pan, Rangeet and Rajan, Hridesh},
title = {A Comprehensive Study on Deep Learning Bug Characteristics},
year = {2019},
isbn = {9781450355728},
publisher = {Association for Computing Machinery},
address = {New York, NY, USA},
url = {https://doi.org/10.1145/3338906.3338955},
doi = {10.1145/3338906.3338955},
booktitle = {Proceedings of the 2019 27th ACM Joint Meeting on European Software Engineering Conference and Symposium on the Foundations of Software Engineering},
pages = {510–520},
numpages = {11},
location = {Tallinn, Estonia},
series = {ESEC/FSE 2019}
}

@inproceedings{Islam2020ICSE,
author = {Islam, Md Johirul and Pan, Rangeet and Nguyen, Giang and Rajan, Hridesh},
title = {Repairing Deep Neural Networks: Fix Patterns and Challenges},
year = {2020},
isbn = {9781450371216},
publisher = {Association for Computing Machinery},
address = {New York, NY, USA},
url = {https://doi.org/10.1145/3377811.3380378},
doi = {10.1145/3377811.3380378},
booktitle = {Proceedings of the ACM/IEEE 42nd International Conference on Software Engineering},
pages = {1135–1146},
numpages = {12},
location = {Seoul, South Korea},
series = {ICSE '20}
}

@misc{stackoverflow:20076195,
  author       = {{Stack Overflow user}},
  title        = {What is the most efficient way of counting occurrences in pandas?},
  year         = {2013},
  url          = {https://stackoverflow.com/questions/20076195/what-is-the-most-efficient-way-of-counting-occurrences-in-pandas},
  note         = {Accessed: 2025-07-19},
  howpublished = {\url{https://stackoverflow.com/questions/20076195/what-is-the-most-efficient-way-of-counting-occurrences-in-pandas}}
}

@misc{meta2024llama3,
  author       = {Meta AI},
  title        = {LLaMA 3: Open and Efficient Foundation Language Models},
  year         = {2024},
  url          = {https://ai.meta.com/llama/},
  note         = {Accessed: 2025-07-19}
}

@article{yang2025qwen3,
  title={Qwen3 technical report},
  author={Yang, An and Li, Anfeng and Yang, Baosong and Zhang, Beichen and Hui, Binyuan and Zheng, Bo and Yu, Bowen and Gao, Chang and Huang, Chengen and Lv, Chenxu and others},
  journal={arXiv preprint arXiv:2505.09388},
  year={2025}
}

@article{abdin2024phi,
  title={Phi-4 technical report},
  author={Abdin, Marah and Aneja, Jyoti and Behl, Harkirat and Bubeck, S{\'e}bastien and Eldan, Ronen and Gunasekar, Suriya and Harrison, Michael and Hewett, Russell J and Javaheripi, Mojan and Kauffmann, Piero and others},
  journal={arXiv preprint arXiv:2412.08905},
  year={2024}
}

@misc{StackExchangeQuery,
author = {Stack Exchange},
title = {DS Patch Gen Query},
howpublished = {\url{https://data.stackexchange.com/stackoverflow/query/1907457/ds-patch-gen}},
year = {2025},
note = "[Online; accessed July-2025]"
}

@inproceedings{le-cong-etal-2025-llms,
    title = "Can {LLM}s Reason About Program Semantics? A Comprehensive Evaluation of {LLM}s on Formal Specification Inference",
    author = "Le-Cong, Thanh  and
      Le, Bach  and
      Murray, Toby",
    editor = "Che, Wanxiang  and
      Nabende, Joyce  and
      Shutova, Ekaterina  and
      Pilehvar, Mohammad Taher",
    booktitle = "Proceedings of the 63rd Annual Meeting of the Association for Computational Linguistics (Volume 1: Long Papers)",
    month = jul,
    year = "2025",
    address = "Vienna, Austria",
    publisher = "Association for Computational Linguistics",
    url = "https://aclanthology.org/2025.acl-long.1068/",
    doi = "10.18653/v1/2025.acl-long.1068",
    pages = "21991--22014",
    ISBN = "979-8-89176-251-0"
}

@article{levenshtein1966binary,
  title={Binary codes capable of correcting deletions, insertions, and reversals},
  author={Levenshtein, Vladimir I.},
  journal={Soviet Physics Doklady},
  volume={10},
  number={8},
  pages={707--710},
  year={1966}
}

@inproceedings{10.1145/3180155.3180206,
author = {Tsantalis, Nikolaos and Mansouri, Matin and Eshkevari, Laleh M. and Mazinanian, Davood and Dig, Danny},
title = {Accurate and efficient refactoring detection in commit history},
year = {2018},
isbn = {9781450356381},
publisher = {Association for Computing Machinery},
address = {New York, NY, USA},
url = {https://doi.org/10.1145/3180155.3180206},
doi = {10.1145/3180155.3180206},
booktitle = {Proceedings of the 40th International Conference on Software Engineering},
pages = {483–494},
numpages = {12},
location = {Gothenburg, Sweden},
series = {ICSE '18}
}

@article{wei2022chain,
  title={Chain-of-thought prompting elicits reasoning in large language models},
  author={Wei, Jason and Wang, Xuezhi and Schuurmans, Dale and Bosma, Maarten and Xia, Fei and Chi, Ed and Le, Quoc V and Zhou, Denny and others},
  journal={Advances in neural information processing systems},
  volume={35},
  pages={24824--24837},
  year={2022}
}

@inproceedings{10.1145/3338906.3338955,
author = {Islam, Md Johirul and Nguyen, Giang and Pan, Rangeet and Rajan, Hridesh},
title = {A comprehensive study on deep learning bug characteristics},
year = {2019},
isbn = {9781450355728},
publisher = {Association for Computing Machinery},
address = {New York, NY, USA},
url = {https://doi.org/10.1145/3338906.3338955},
doi = {10.1145/3338906.3338955},
booktitle = {Proceedings of the 2019 27th ACM Joint Meeting on European Software Engineering Conference and Symposium on the Foundations of Software Engineering},
pages = {510–520},
numpages = {11},
location = {Tallinn, Estonia},
series = {ESEC/FSE 2019}
}

\end{document}